\RequirePackage{silence}
\documentclass[fleqn,usenatbib,useAMS]{mnras} %Class file altered on lines 116, 119, 122 and 123.
\usepackage[utf8]{inputenc}
\DeclareUnicodeCharacter{2212}{-}
\usepackage{graphicx}
\usepackage{threeparttable}
\usepackage{scrextend}
\usepackage{tablefootnote}
\usepackage{amsmath}

\usepackage{amsfonts}
\usepackage{float}
\usepackage{bm}
\usepackage{ae,aecompl}
\usepackage{array}
\usepackage{soul}
\usepackage{mathtools}
\usepackage{multirow}
\usepackage{bigdelim}
\newcolumntype{_}{>{\global\let\currentrowstyle\relax}}
\newcolumntype{^}{>{\currentrowstyle}}

\newcommand{\appropto}{\mathrel{\vcenter{
			\offinterlineskip\halign{\hfil$##$\cr %The $##$ is not a mistake!
				\propto\cr\noalign{\kern2pt}\sim\cr\noalign{\kern-2pt}}}}}

\title[BTFR in the varying IMF]{The effect of the environment-dependent stellar initial mass function on the baryonic Tully Fisher relation} %Use [Short title]{Full title} if the two are inequivalent, former should be <=45 characters unless you want to explicitly make it multi-line.

\author[Zonoozi et al.]
{Akram Hasani Zonoozi$^{1,2}$\thanks{
E-mail:  \mbox{a.hasani@iasbs.ac.ir; e.h.zonoozi@gmail.com} (AHZ)}, Hosein Haghi$^{1,2,3}$
, Pavel Kroupa$^{2, 4}$\thanks{
E-mail:  \mbox{pkroupa@uni-bonn.de} (PK)}, Sara Yousefizadeh$^{1}$ \newauthor 
Zhiqiang Yan$^{5,6}$, Tereza Jerabkova$^{7}$, and Eda Gjergo$^{5,6}$\\
$^{1}$Department of Physics, Institute for Advanced Studies in Basic Sciences (IASBS), PO Box 11365-9161, Zanjan, Iran\\
$^{2}$Helmholtz-Institut f\"ur Strahlen-und Kernphysik (HISKP), Universit\"at Bonn, Nussallee 14-16, D-53115 Bonn, Germany\\
$^{3}$School of Astronomy, Institute for Research in Fundamental Sciences (IPM), PO Box 19395 − 5531, Tehran, Iran\\
$^{4}$Charles University in Prague, Faculty of Mathematics and Physics, Astronomical Institute, V Hole\v{s}ovi\v{c}k\'ach 2, CZ-180 00 \\
Praha 8, Czech Republic\\
$^{5}$Nanjing University, School of Astronomy and Space Science, Nanjing 210093, People’s Republic of China\\
$^{6}$Nanjing University, Key Laboratory of Modern Astronomy and Astrophysics, Ministry of Education, Nanjing 210093,\\
People’s Republic of China\\
$^{7}$Department of Theoretical Physics and Astrophysics, Faculty of Science, Masaryk University, Kotl\'a\v{r}sk\'a 2, Brno 611 37, Czech Republic\\
}

\pubyear{2022}
\begin{document}
\label{firstpage}
\pagerange{\pageref{firstpage}--\pageref{lastpage}}

\maketitle

\begin{abstract}

 We investigate the impact of an environment-dependent galaxy-wide stellar initial mass function (gwIMF) on the baryonic Tully-Fisher relation (BTFR). The integrated galaxy-wide IMF (IGIMF) theory, which incorporates variations in stellar populations due to star formation history (SFH) and metallicity, provides a more accurate framework for understanding systematic deviations in galaxy scaling relations than that given by an invariant gwIMF. By considering how the mass-to-light ratio of the stellar population is influenced by metallicity and SFH, we show that high-mass galaxies have their masses in stars and remnants underestimated under the assumption of a constant mass-to-light ratio. In contrast, low-mass, gas-dominated galaxies are less affected. Our results suggest that the discrepancies between the true and observed BTFR are primarily driven by the evolving nature of the stellar IMF, particularly in galaxies with slowly declining SFHs. The IGIMF theory offers a solution to the observed offsets in the BTFR, especially for high-mass galaxies, where the rotational velocities are higher than predicted by MOND. We conclude that incorporating the IGIMF provides a more accurate description of galaxy dynamics, revealing the importance of stellar population characteristics in refining our understanding of the baryonic mass-velocity relationship. This study underscores the necessity of accounting for the variation of the gwIMF when interpreting the BTFR, particularly in the context of alternative gravitational theories like MOND. 

\end{abstract}


\section{Introduction}
\label{sec:Introduction}

The phenomenon of gravitation remains poorly understood. In the strong-field regime it is excellently described by the geometric distortion of space-time through matter \citep{Einstein1916, Kramer+21}. However, the fundamental nature of gravity is still debated, with alternative perspectives suggesting it may be an emergent phenomenon arising from spacetime information content \citep{Verlinde2017}, or a manifestation of the wave properties of matter \citep{Stadtler2021, Zhang2022}. In the non-relativistic limit, general relativity reduces to Newtonian gravity, which has been remarkably successful within the Solar System. However, on galactic and cosmological scales, observations such as flat galaxy rotation curves and gravitational lensing require either a significant dark matter component or modifications to the laws of gravity.  The standard approach within the current $\Lambda$CDM cosmological model posits the existence of non-baryonic dark matter particles, which have not yet been directly detected despite extensive searches \citep{Kroupa12, Kroupa15, Roshan+21, Asencio+22, Kroupa2023, Oehm2024}. 

Another theoretical approach is to argue that the law, which was formulated by Newton in the 17th century on the basis of observational data confined solely to the Solar system (in 1916 this had not changed), amends when the gravitational acceleration is smaller than Milgrom's constant $a_0\approx 3.8\,$pc/Myr$^2$ \citep{Begeman1991}. This concept allowed \cite{BekensteinMilgrom1984} and \cite{Milgrom2010} to develop new non-relativistic Lagrangians, the extremisation of which lead to similar generalised Poisson equations. One prediction of this theory \citep{Milgrom1983b}  is that isolated galaxies must have a circular velocity in the asymptotically flat part of their rotation curve which is
\begin{equation}
    V_{\rm f} = \left(G\,a_0\, M_{\rm b} \right)^{1/4}, %\approx m_{\rm Mil}\,\left(M_{\rm b} \right)^{1/4},
    \label{eq:BTFR0}
\end{equation}
where $M_{\rm b}$ is the baryonic mass of the galaxy (comprised of all stars, $M_*$, stellar remnants, $M_{\rm rem}$, and gas, $M_{\rm g}$), $G\approx 0.0045\,{\rm pc}^3\,M_\odot^{-1}\,{\rm Myr}^{-2}$ and $V_{\rm f}$ in units of pc/Myr ($\approx \,$km/s).

It is therefore of significant importance to test if observed galaxies over all accessible masses comply with the prediction (Eq.~\ref{eq:BTFR0}) or if deviations would lead to the falsification of the theory. While MOND has had some empirical successes, especially on galaxy and star cluster scales \citep{FamaeyMcGaugh2012, BanikZhao2021, Kroupa2022, Kroupa2024a}, it remains a subject of ongoing debate within the astrophysical community, and is not a mainstream consensus theory. Our study aims to test this prediction empirically using observed galaxy data.

A correlation between the luminosity of a galaxy (which depends on the stellar mass of the galaxy $M_*$) and the width of its global HI profile (which is an indicator of the circular velocity) was discovered by \cite{Tully1977} and is known as the Tully-Fisher (TF) relation. The  TF  relation, however, breaks down for velocities smaller than about 100 km/s \citep{McGaugh2000}, a regime dominated by gas-rich dwarf galaxies with their gas mass ($M_{\rm g}$) being comparable to or even larger than  $M_*$.  Replacement of $M_*$ by $M_{\rm b}$ leads to the verification of Milgrom's prediction (Eq.~\ref{eq:BTFR0}) over 5~orders of magnitude in baryonic mass \citep{McGaugh2000, McGaugh2005, Lelli2016}. The empirical form of the relationship has been named by \cite{McGaugh2000} as the baryonic Tully–Fisher relation (BTFR). Rotationally and pressure-supported galaxies have been shown to obey Milgrom's prediction precisely and accurately for $M_{\rm b}<10^{11}\,M_\odot$ \citep{McGaugh2015, Ponomareva2018, McGaugh+21}.

In massive disk galaxies, $M_{\rm g} \approx 0.1\, M_*$, such that gas masses play a subordinate role in defining the gravitational potential which is thus defined, in MOND, predominantly by the stellar population. It is the purpose of this contribution to investigate how systematic differences in the stellar populations of galaxies lead to apparent systematic deviations from Milgrom's prediction (Eq.~\ref{eq:BTFR0}) if the data reduction does not take into account these systematic population differences. In the analysis by \citet{Lelli2016}, the stellar masses of galaxies are estimated using luminosity at 3.6 microns, assuming the stellar populations all have the same mass-to-light ratio in Solar units, $(M_{\rm pop}/L_{\rm [3.6]} = 0.5)$\footnote{ Throughout this paper, the unit of the mass-to-light ratio is consistently expressed as $M_{\odot}/L_{\odot}$.}. The value of the $M_{\rm pop}/L_{\rm [3.6]}$ ratio  which represents the mass-to-light ratio of the stellar population including remnants, depends on the galaxy-wide stellar initial mass function (gwIMF) and on the star-formation history (SFH) of the galaxy. The fastest rotator in their sample requires $M_{\rm pop}=M_* + M_{\rm rem}$ to be greater by a factor of a few than estimated using a canonical stellar initial mass function (canIMF) to match the high-mass BTFR and therefore Eq.~\ref{eq:BTFR0}.

The variation of the gwIMF with environmental factors such as metallicity and star formation rate (SFR) density has been an important aspect of recent studies. Observational evidence has shown that the stellar IMF is not invariant across different environments, with notable variations observed at both the high and low-mass ends. \citet{Jerabkova+18} and \citet{Yan2017} highlighted that the gwIMF can exhibit significant variations with the SFR, becoming top-heavy in high-SFR environments and top-light in low-SFR dwarf galaxies as shown by observational surveys  \citep{Lee09, Gunawardhana11, Zhang+18, Jerabkova+18}. These variations are also observed for low-mass stars, with the gwIMF becoming bottom-light in metal-poor environments \citep{ Yan+20, Li2023}, and bottom-heavy in galaxies with super-solar metallicities \citep{Martin-Navarro2019, Smith2020, vanDokkum2021}. 
Furthermore, \citet{Yan2024} discussed the impact of the low-mass end of the stellar IMF, which, due to its dominance in stellar mass, has profound effects on stellar mass estimations and galaxy evolution. They noted that the stellar IMF of low-mass stars tends to be more dependent on metallicity than on other parameters such as SFR \citep{Martin-Navarro2015, Parikh2018, Zhou2019}. In addition, the environmental-dependent gwIMF has proven successful in explaining the dynamics of the ultra-diffuse galaxy Dragonfly 44 \citep{Haghi2019}. These findings provide compelling motivation for studying the impact of systematic changes in gwIMF on BTFR, which this study aims to explore.

This study examines how the systematic variation of the gwIMF with metallicity and star formation rate (SFR) might affect the BTFR. We test whether the integrated galaxy-wide IMF (IGIMF) theory \citep{Kroupa03, Kroupa13, Jerabkova+18, Yan+21, Kroupa2024b} can account for the observed deviations from the BTFR, and address the expected systematic deviations as a function of the baryonic mass of star-forming galaxies. By assuming both Milgrom's prediction (Eq.~\ref{eq:BTFR0}) and the IGIMF theory are correct, we calculate the luminosity of the model galaxy for a given \(V_{\rm f}\) and SFH. The observer will use this luminosity to calculate the stellar mass assuming an invariant canonical stellar IMF (i.e. a given constant mass-to-light ratio $M_{\rm pop}/L_{\rm [3.6]} = 0.5$), and the so-calculated baryonic mass will be offset from the true BTFR because the true $M_{\rm pop}/L_{\rm [3.6]}$ can be larger or smaller than 0.5 depending on the SFH of the galaxy.

In Sec.~\ref{sec:method}, the gwIMF as applied here is documented, the method for computing the gwIMF is explained, the stellar evolution used, and the properties of the stellar remnants are described. The  $M_{\rm pop}/L_{\rm [3.6]}$ ratio of galaxies with different present-day stellar masses in the context of the invariant canIMF and IGIMF is compared in  Sect.~\ref{sec:MLratio}. The effect of systematically varying gwIMF on the BTFR is explored in Sect.~\ref{sec:BTFR}. Sec.~\ref{sec:Conclusion} contains the conclusions.

\section{Methods}
\label{sec:method}

A grid of model galaxies is generated following the procedure described in Sec.~\ref{sec:procedure} according to which stellar and remnant populations are calculated assuming an IMF (Sec.~\ref{sec:IMF}), stellar evolution (Sec.~\ref{sec:stevol}) and SFH (Sec.~\ref{sec:SFHs}). Throughout this work we reffer to the stellar IMF as the IMF of all stars formed in one molecular cloud clump. The gwIMF is the composite IMF of all stars that form in all embedded clusters (i.e., molecular cloud clumps) in the galaxy. The gwIMF can be asuumed to be the invariant canonical IMF (canIMF, Sec. 2.2.1), or be given by the IGIMF (Sec. 2.2.2).

\subsection{Procedure}
\label{sec:procedure}

The aim is to quantify where a model galaxy lies relative to the theoretical BTFR (Eq.~\ref{eq:BTFR0}). For each computed galaxy model, the total mass in all stars formed is $M_{\rm tot}$ (Eq.~\ref{eq:Mtot} below). At the present-day, the baryonic mass of a model galaxy is 
\begin{equation}
    M_{\rm b} = M_* + M_{\rm rem} + M_{\rm g}=M_{\rm pop}+ M_{\rm g},
\label{eq:Mb}
\end{equation}
where $M_{*}$, $M_{\rm g}$, and $M_{\mathrm{rem}}$ are the mass of the living stars, the gas disk,  and stellar remnants, respectively. 

Using the Fortran code \emph{SPS-VarIMF}\footnote{https://github.com/ahzonoozi/SPS-VarIMF} \citep{Zonoozi2025} a grid of galaxy models is computed with the {\it idealized galaxy models} being based on the invariant canonical IMF and the {\it realistic galaxy models} being based on the IGIMF theory. Unless stated otherwise, the galaxies are assumed to start to form $12.5\,$Gyr ago and continue to grow until the present time. At each  $\delta t=10 $ Myr time-step a new stellar population is added, while the previously formed stellar populations evolve, leaving giants and remnants in the galaxy and longer-lived stars.

The initial metallicity of the gas in the galaxy is assumed to be $Z_*=0.0002$.  As galaxies evolve, they become self-enriched through the evolution of massive stars, which inject metals into the interstellar medium via stellar winds and supernova explosions. This enriched gas then forms a new generation of stars, raising the average stellar metallicity.
In this model, a closed-box approach is adopted with a time-integrated star formation efficiency $f_{\rm st}=M_{\rm tot}/M_{\rm g}=$0.3, where $M_{\rm tot}$ is the total stellar mass formed over the age of the galaxy. The star formation efficiency represents the ratio of the total stellar mass formed throughout the galaxy's entire history to the total mass of the gas involved.

For a realistic model, we expect the final mass and metallicity of simulated galaxies to align with the observed mass-metallicity relation \citep{Ma+16}. This relation suggests that more massive galaxies tend to have higher metallicities, as they have undergone longer and more intense star formation histories, enriching their interstellar medium with metals,
\begin{equation} 
\begin{split}
    {\rm log}_{10}(Z_*/Z_{\odot}) = 0.40\left[{\rm log}_{10}(M_*/M_{\odot})-10\right] \\ 
    + \; 0.67\, e^{-0.50z} - 1.04,
    \label{eq:MZrel}
\end{split}
\end{equation}
where $M_*$ is the final stellar mass of the model and we assume redshift $z=0$.
Eq.~\ref{eq:MZrel} is a linear fit to the data shown in fig.~2 of \citet{Ma+16}.
In stellar population synthesis (SPS) models, the final metallicity of a galaxy is determined by several key factors \citep{Gjergo2023}, including the SFH, the chemical enrichment process, and the feedback mechanisms that regulate gas and star formation. By making specific assumptions about these factors such as the initial metallicity, the rate of gas infall, and the star formation efficiency ($f_{\rm st}$)\footnote{Note that the definition of "star formation efficiency" used here differs from that used in \citet{Gjergo2023}.}, one can adjust the galaxy's final metallicity.

However, our models do not include gas flows, and the same star formation efficiency is assumed for all models. Nevertheless, as a direct consequence of the IGIMF theory, more massive galaxies will have higher final metallicities. This is because in these galaxies, which have larger star formation rates (SFR), the galaxy wide IMF (gwIMF) becomes more top-heavy, leading to the formation of a relatively greater number of massive stars. As these stars evolve and die, they enrich the surrounding interstellar medium by injecting a larger quantity of metals, thereby increasing the gas-phase metallicity.

The gas mass contribution to $M_{\rm b}$ is computed based on the methodology outlined by \citet{Lelli2016}. The detected gas mass of the galaxy depends on its $[3.6]$-band luminosity as described in Eq. 4 of \citet{Lelli2016}, i.e.
\begin{equation}
    {M_{\rm HI}(L_{\rm [3.6]}) \over M_\odot} = 10^{3.9}\, \left( {L_{\rm [3.6]}\over L_{\odot,{\rm [3.6]}} } \right)^{0.54},
    \label{eq:Mg}
\end{equation}
where \( M_{\rm HI}\) represents the HI mass, and the total gas mass is given by \(M_{\rm g} = 1.33 M_{\rm HI}\), with the factor 1.33 accounting for the contribution of helium. However, we note that this approach relies on a single measurement of \(L_{\rm [3.6]}\) to estimate both stellar and gas masses, which introduces uncertainties, particularly given the variable gas-to-stellar mass ratio. Additionally, gas mass conversion factors are known to have significant uncertainties and dependencies on metallicity \citep{Bisbas2025}, which are not explicitly accounted for in this method.

The code \emph{SPS-VarIMF} is employed here to calculate the stellar populations and their luminosities as a function of the SFR and metallicity, which is based on the first version having been published in an application of the IGIMF theory to the evolution of the Milky Way as a former chain galaxy \citep{Zonoozi2019}. See also \citet{Yan2017}, \cite{Yan+21} and \citet{Haslbauer2024} for the python module {\it GalIMF} and {\it PhotGalIMF} developed for a similar purpose.

\subsection{The stellar and galaxy-wide IMF}
\label{sec:IMF}
The number of stars born together in one embedded cluster (i.e. in a molecular cloud clump) in the initial-stellar-mass interval $m$ to $m+dm$, is $dN=\xi_*(m)dm$, where $\xi_*(m)\propto m^{-\alpha_i}$ is the stellar IMF. The stellar IMF is described as a multipower-law function with different indices for various mass ranges, as proposed in  earlier analyses of resolved stellar populations \citep{KTG93, Kroupa02a, Bastian10, Offner14, Hopkins18}. Specifically, for the stellar IMF, the function follows:
\begin{equation} \label{CanIMF}
\xi_\star( m)= k_\star\left\{
      \begin{array}{ll}
\,2  m^{-\alpha_1},      & 0.08  M_\odot\leq m < 0.5  M_\odot,  \\
\,  m^{-\alpha_2},      & 0.5  M_\odot\leq  m < 1.0  M_\odot,\\
\,  m^{-\alpha_3},      & 1.0  M_\odot\leq  m <  m_{\rm max}.\\
       \end{array}
        \right. 
\end{equation}
The galaxy models are calculated under two different assumptions for the gwIMF:

\subsubsection{Galaxy models with the invariant canonical gwIMF: idealized models}
\label{sec:ideal}

The assumed-invariant canonical two-part power-law IMF (the canIMF) has $\alpha_1=1.3$ with $\alpha_2=\alpha_3=2.3$ being the Massey-Salpeter power-law index \citep{Kroupa01, Kroupa13, Kroupa2024a}.  The stellar population and the star formation activity in the solar neighborhood characterize it. 

The canonical galaxy models assume the galaxy-wide IMF (gwIMF) to be identical in shape to the invariant canIMF (Eq. \ref{CanIMF}) and to have a fixed stellar mass range of \( 0.08 \leq m/M_\odot \leq 150 \) in all cases. As a result, these models can include a fraction of massive stars, even for dwarf galaxies, disfavored by observations \citep{Lee09}. Such models are intended to serve as benchmarks for comparison against more realistic galaxy models computed using the IGIMF theory which does not allow fraction of stars to be present. In these idealized models, characterized by the  canIMF, the mass-to-light ratio depends solely on the metallicity of the stellar population (\(Z_*\)) and the shape of the SFH.
The mass of the most massive star that forms in an embedded cluster, $m_{\rm max}$, is related to the mass of the cluster through a specific relation and normalization of the mass function of the embedded cluster (see, e.g., \citealt{Yan2023, Kroupa2024b}).

\subsubsection{Galaxy models with the IGIMF theory: realistic models}
\label{sec:realistic}

Research on the soloar neighborhood, Galactic bulge,  star-burst clusters, ultra-compact dwarf galaxies, and globular clusters (\citealt{Kroupa02a,Dab+10,Dab+12,Marks12, BK12, Zonoozi16, Haghi17, Jerabkova+17, Hopkins18} and references therein,\citealt{Li2023}) has uncovered a dependency of the power-law indices, $\alpha_i$, in Eq. \ref{CanIMF} on the physical conditions of the star-forming clouds, such that they are functions of the metallicity, $Z_*$, and volume mass density of the molecular cloud clump in which the embedded star cluster forms. Here the most-up-to-date formulation of these functions, obtained by including constraints from star-forming dwarf galaxies, ultra-diffuse and massive elliptical galaxies \citep{Jerabkova+18, Yan+20, Yan+21} is adopted. Specifically, eq.~1--7 in \cite{Yan+21} are applied to define the IMF in an embedded cluster with stellar mass $M_{\rm ecl}$ and metallicity $Z_*$.  
The stellar mass range extends from $m=0.08\,M_\odot$ to $m_{\rm max}\le 150\,M_\odot$, where $m_{\rm max}$ is a function of $M_{\rm ecl}$ as a result of self-regulated embedded-cluster formation \citep{Yan2023, Chavez2025}. This function is obtained by solving eq.~9 and~10 in \cite{Yan+21}.

In the IGIMF theory, the stellar IMF depends on the mass, $M_{\rm ecl}$, and metallicity, $Z_*$, of the embedded cluster via  
\begin{equation} \label{alpha_z1}
\begin{split}
\alpha_1(Z_*)=1.3 +\Delta \alpha . (Z_*-Z_{\odot}), \\
\alpha_2(Z_*)=2.3 +\Delta \alpha . (Z_*-Z_{\odot}),\\ 
\alpha_3=\alpha_3(Z_*, M_{\rm ecl}),\\
\end{split}
\end{equation} 
where $\Delta \alpha=63$ was obtained by \cite{Yan+20} and \cite{Yan+21}. $Z_*$ and $Z_{\odot}= 0.0142$ \citep{Asplund2009} are the mean stellar metal-mass fractions of the system and the Sun, respectively. That is, $\alpha_1$ and $\alpha_2$ are functions of metallicity, while $\alpha_3$ is a function of both metallicity and mass in stars of the embedded cluster that forms in a molecular cloud clump. For a detailed formulation of $\alpha_3$, see Eqs. 6 and 9 of \citet{Jerabkova+18}.

The stellar IMF-relevant upper limit of stellar masses of $150\,M_\odot$ adopted here stems from star-counts in young stellar populations with more massive stars emerging from mergers in the binary-rich birth populations (e.g. \citealt{Banerjee+12a, Wang+20} and references therein; see also \citealt{Oh18}). 
More recent observational support for the variation of the stellar IMF with metallicity and density has come forth in direct star counts in the R136 star-burst in the Large Magellanic Cloud \citep{Schneider+18} and in the Magellanic Bridge cluster NGC~796 \citep{Kalari+18}, through observations of tidal disruption events \citep{Bortolas22} and chemical enrichment in galactic nuclei \citep{Toyouchi2022}.

The gwIMF is only equal in shape to the stellar IMF if the IMF is interpreted to be an invariant probability density distribution function \citep{Kroupa13, KJ18, KJ21, Kroupa2024b}. If, instead, the stellar IMF varies with metallicity and density and, as assumed here, if it is an optimally-sampled distribution function (i.e. star-formation is highly self regulated) then the gwIMF depends on $Z_*$ and $SFR$ of the galaxy, and can be computed using the IGIMF theory \citep{Yan2017, Jerabkova+18, Yan2023}. According to the fundamental assumption of the IGIMF theory, the stellar population of a galaxy is built up by stars forming in embedded clusters \citep{Kroupa03, DK22} such that the gwIMF is the integral over all embedded clusters that form at a given time. The SFR of a galaxy thus determines the distribution of densities reached in the inter-stellar medium of a galaxy and thus the most massive embedded cluster that can form at a given time. 

Since the shape of the stellar IMF depends on the gas cloud-clump density and thus on the mass of an embedded cluster, it follows that the gwIMF varies. It is top-light (lacking massive stars) at small SFRs ($SFR<10^{-2}\,M_\odot$/yr) and becomes top-heavy (over surplus of massive stars) at high SFRs ($SFR>10\,M_\odot$/yr). For a given SFR, the gwIMF is more bottom-light (lacking low-mass stars) and more top-heavy at smaller sub-solar metallicity, and it is more bottom-heavy and more top-light with increasing $Z_*$ \citep{Jerabkova+18, Yan+21}. The most up-to-date formulation of the IGIMF theory (eq.~8--12 in \citealt{Yan+21}) is adopted here. We emphasize that, as shown in \cite{Yan+21}, this formulation for the first time brought the formation time-scales of elliptical galaxies, as deduced from their metallicities and alpha-element abundances, into agreement with those deduced from stellar-population modelling. 

The formulation of the variation of the stellar IMF with metallicity and density (Eq. \ref{alpha_z1}) is not ad hoc, but has been developed over time by ensuring the calculated embedded clusters evolve to systems that are consistent with observed clusters and ultra-compact dwarf galaxies in the Milky Way \citep{Mahani2021}, subject to additional constraints from ultra-diffuse and elliptical galaxies. It is thought that this variation realistically captures the variation of stellar populations across cosmic time \citep{Chruslinska+20} as it also explains the properties of super-massive black holes and their hosting galaxies \citep{Kroupa+20}. 

Observational support for the IGIMF theory comes from the deficit of H${\alpha}$ emission in star-forming dwarf disk galaxies with small SFRs \citep{Lee09}, a lack of massive stars in the nearby dwarf galaxy DDO~154 \citep{Watts18}, the radial H${\alpha}$ cutoff in UV-extended galactic disks \citep{Pflamm-Altenburg2008}, top-heavy gwIMFs observed in disk galaxies with high SFRs \citep{Gunawardhana11} and top-heavy gwIMFs indicated in C-isotopologue ratios in distant starburst galaxies \citep{Zhang+18},  ultra-violet spectra of extreme starbursts 
\citep{Senchyna+21} and observations of Wolf-Rayet galaxies \citep{Liang+21}. A similar trend has been found in early-type galaxies with the gwIMF being more bottom-heavy at higher metallicity (e.g. \citealt{Rosani+18, Zhou2019}).

\subsection{Stellar evolution and dark remnants}
\label{sec:stevol}

We used the latest stellar evolution tracks from the PARSEC database to calculate the properties of evolving stellar populations in our galaxy models. The PARSEC database is a comprehensive set of theoretical models developed by the Padova and Trieste groups \citep{Bressan2012, Marigo2017}, which simulate the evolution of stars across a wide range of masses, metallicities (0.0002 to 0.06), and ages from the pre-main sequence to the end of the star's life.  The models cover different stellar phases, including the main sequence, giant phases, and final stages of stellar evolution, such as the formation of white dwarfs, neutron stars, or black holes, depending on the progenitor mass and metallicity. All dark remnants remain in the galaxy models.  To assign remnant masses to dead stars, we use the metallicity-dependent initial-final mass relation from \cite{Spera2015}. This prescription accounts for the impact of stellar metallicity on mass loss during stellar evolution, leading to more massive remnants in lower-metallicity environments.

\subsection{Star formation histories}
\label{sec:SFHs}

The galaxy models start forming at time $t=0$, and star formation continues for a time $t_{\rm sf}$. The age of galaxies is assumed to be $t_{\rm sf}=12.5\,$Gyr.
For each SFH, the total stellar mass formed is
\begin{equation}
    M_{\rm tot} = \int_0^{t_{\rm sf}}\, SFR(t) \, dt,
    \label{eq:Mtot}
\end{equation}
where, $SFR(t)$ is assumed to be constant, as nearby disk galaxies exhibit a nearly constant SFR over time \citep{ Kroupa+20b}, or slowly declining, 
\begin{equation}
\psi(t)= \psi_0  e^{-t/ \tau}, 
\label{eq:DecSFR}
\end{equation}
with an exponential timescale of $\tau=10 \rm ~Gyr$. Here,  $t$ is the age of the galaxy and $\psi_0$ is the SFR at the beginning of the burst.
These galaxies appear to be surrounded by reservoirs of gas that replenish the ISM, helping to sustain star formation. The cold gas required for star formation is continuously supplied through accretion from the intergalactic medium (IGM), while stellar feedback—via winds, supernovae, and other processes—regulates the ISM, preventing star formation from becoming too intense or shutting down entirely. Given that the present-day SFRs of galaxies in the Local Cosmological Volume is comparable to their mean SFR over cosmic time, assuming a constant SFR is well justified for these galaxies \citep{Kroupa+20b}. 

The empirical scaling relations used in this study, such as the mass-metallicity relation (Eq. \ref{eq:MZrel}) from \citet{Ma+16} and the BTFR, are primarily derived from observational samples of star-forming disk galaxies. These galaxies, often on the main sequence, exhibit sustained star formation over cosmic time, making them suitable for comparison with our models that assume continuous or slowly declining SFHs.

\section{The mass-to-light ratio in the IGIMF and canonical IMF context}
\label{sec:MLratio}

The mass-to-light ($M/L$) ratio is essential for estimating the total mass content of a galaxy. In observational studies of galaxies, it is often assumed that the $M/L$ ratio is constant or varies predictably with galaxy type or other parameters \citep{Schombert+19}. This assumption simplifies the analysis and modeling of galaxy populations, as it provides a straightforward relationship between the mass of a galaxy and its observed luminosity. 
Assuming an invariant canonical stellar IMF, most of the luminosity is contributed by newly formed high-mass stars, while the majority of the mass is concentrated in low-mass stars. However, within the framework of the IGIMF, where the gwIMF shape is influenced by the SFH, the contribution of stellar remnants to the total mass of a galaxy can become substantial, particularly in massive galaxies with a high SFR. Conversely, in very low-mass galaxies with a low SFR, few, if any, high-mass stars are formed, leading to intermediate-mass stars dominating the galaxy's luminosity. 

The mass-to-light ratio of these models depends on $Z_*$, on the shape of the SFH as well as on its amplitude. Note that this variation of the mass-to-light ratio is caused in-part by the stellar IMF power-law-indices $\alpha_1$ and $\alpha_2$ depending on $Z_*$ such that a metal-enriched population will have a larger relative number of M~dwarfs \citep{Yan+21, Yan2024, Haslbauer2024}.
Therefore, a metal-rich galaxy with the same total stellar mass, and with the same mass in remnants, as a metal-poor galaxy, will have a larger mass-to-light ratio.

However, when considering a metallicity-dependent mass for remnants, predicting the mass-to-light ratio of metal-rich versus metal-poor galaxies becomes more complicated, as metal-poor stars leave more massive remnants compared to their metal-rich counterparts. 

\begin{figure}
    \includegraphics[width=\columnwidth]{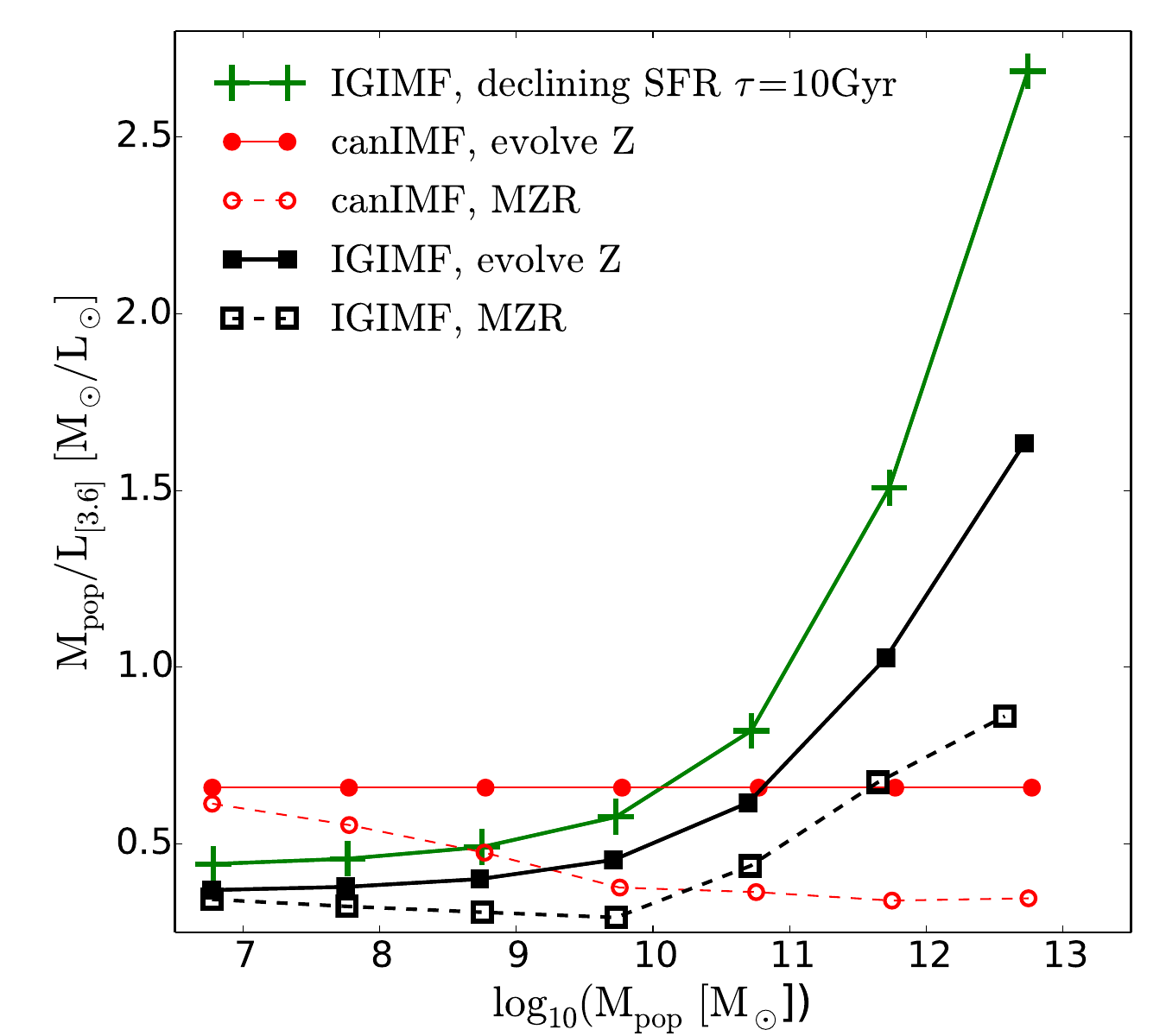}
     \includegraphics[width=\columnwidth]{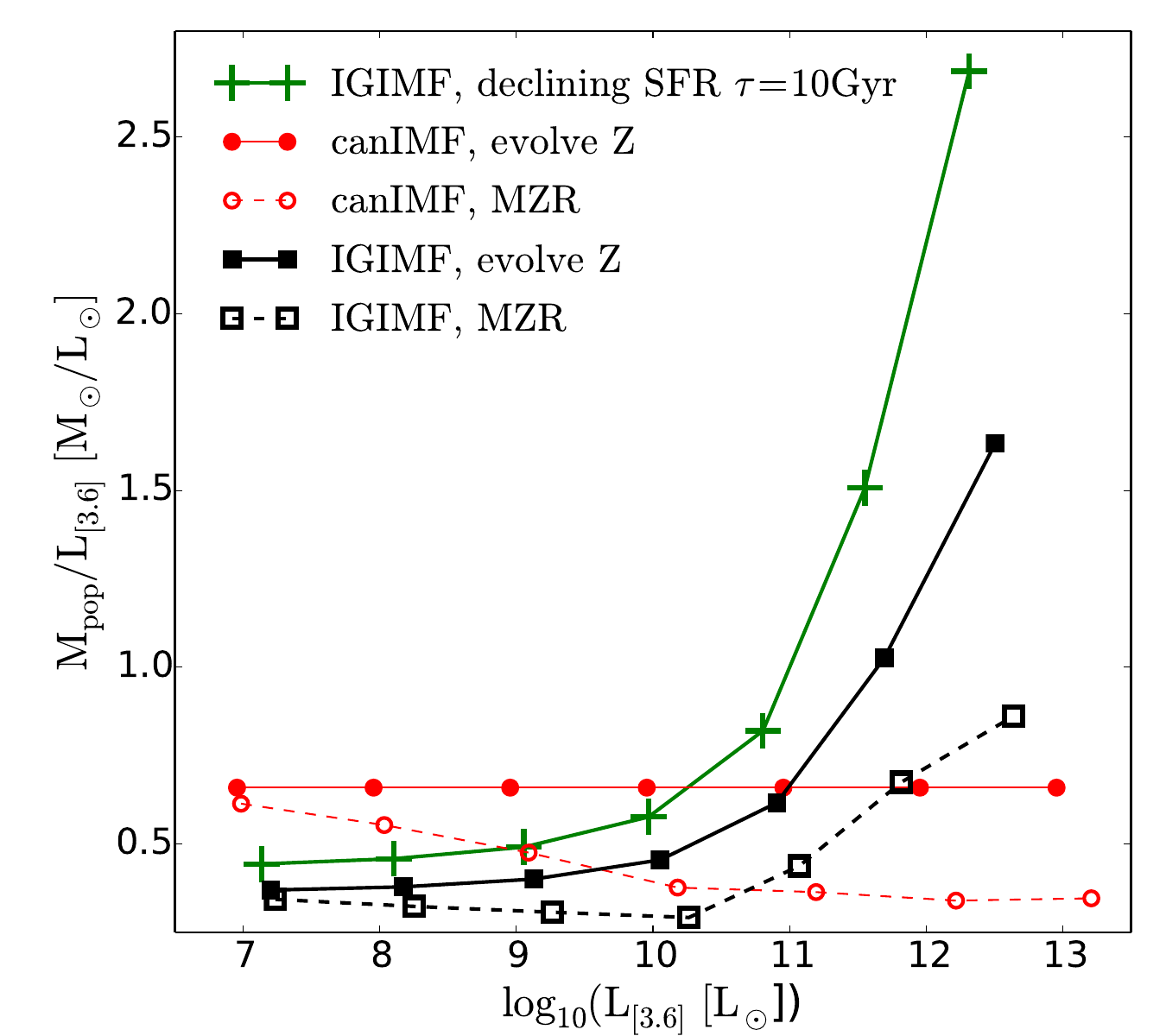}
    \caption{$M_{\rm pop}/L_{[3.6]}$ in dependence on the present-day stellar plus remnant mass ($M_{\rm pop}$ top panel) and luminosity in the [3.6]-band (lower panel) assuming constant and slowly declining SFHs and constant and time-evolving metallicities. Circle symbols represent models with an invariant canonical IMF, while the plus signs correspond to the IGIMF case with a declining SFR, and square symbols represent the IGIMF with a constant SFR. The solid lines illustrate models with a time-evolving metallicity, while the dashed line represents models with a constant metallicity, adopted from the mass-metallicity relation (Eq. \ref{eq:MZrel}). Note that the plotted mass-to-light ratio counts the mass in all stars plus the remnants. }
    \label{figure_ML}
\end{figure}

We define the mass-to-light ratio in the 3.6 $\mu$m photometric band of the stellar population as $M_{\rm pop}/L_{[3.6]}$, where $M_{\rm pop}=M_* + M_{\rm rem}$. In Figure \ref{figure_ML}, the $M_{\rm pop}/L_{[3.6]}$ ratio of computed galaxies are plotted versus their present-day stellar masses (upper panel) and luminosities (lower panel). We compare the $M_{\rm pop}/L_{[3.6]}$ of galaxies with different masses assuming the gwIMF to be the invariant canonical IMF or following the IGIMF theory. For all modelled galaxies a constant SFH is assumed, except for those with a slowly declining SFR. Two models with an invariant IMF are considered: the solid line represents the model that starts with a low metallicity ($Z_*=0.0002$) and undergoes self-enrichment (closed box model), while the dashed line corresponds to a model where each galaxy maintains a constant metallicity determined from the observed mass-metallicity relation (MZR), which is typically based on oxygen abundance (12 + log[O/H]).

Assuming a closed-box model, an invariant canonical IMF as the gwIMF, identical star formation efficiencies, and similar star formation histories (SFHs) for galaxies of different masses results in an identical metallicity enrichment across all galaxies. This assumption contradicts the observed mass-metallicity relation, which shows that more massive galaxies typically exhibit higher metallicities. Under these conditions, the present-day $M_{\rm pop}/L_{[3.6]}$ values of galaxies with different masses will be $M_{\rm pop}/L_{[3.6]} \approx 0.6 M_\odot/L_\odot$ which is plotted as a solid red line in Fig. \ref{figure_ML}.

To ensure consistency with observed metallicities, we calculated a new grid of models based on the invariant canonical IMF (gwIMF=canIMF), with constant metallicities assigned according to the observed present-day mass-metallicity relation (Eq. \ref{eq:MZrel}). As illustrated by the red dashed line in Fig. \ref{figure_ML}, the $M_{\rm pop}/L_{[3.6]}$ decreases as the galaxy mass increases. This trend is primarily driven by the fact that metal-rich stars emit more luminosity in the K band. In addition, metal-poor stars retain more massive remnants, which increase the $M_{\rm pop}/L_{[3.6]}$ ratio in low-mass, metall-poor galaxies.
  
In the IGIMF theory, higher-mass galaxies with larger SFRs undergo a stronger metallicity enrichment, as they form a greater relative number of massive stars that inject more metals into the ISM. As a result, low-mass galaxies retain lower metallicities, while massive galaxies achieve higher metallicities, in agreement with the observed mass-metallicity relation.  
%Note that the SFH plays a significant role in the chemical enrichment of galaxies, particularly in the context of the IGIMF theory. Even with a canonical IMF, the SFH affects average stellar metallicity.
However, adopting different SFHs can significantly influence the chemical enrichment of galaxies, particularly within the framework of the IGIMF. A declining SFR leads to higher average metallicity because more stars form early during rapid chemical enrichment. In the context of the IGIMF, adopting a declining SFR results in older stellar populations being formed under a more top-heavy gwIMF compared to younger populations, which tend to form under a less top-heavy gwIMF. This can lead to a more rapid early enrichment, followed by slower metal production at later times compared to that of the constant SFH models.  However, its cumulative effect over cosmic time (12.5 Gyr) tends to yield almost similar final metallicities for galaxies modeled with constant or mildly declining SFHs ($\tau=10 \rm ~Gyr$), assuming reasonable choices for other parameters. Importantly, the final stellar and gas-phase metallicities also depend on the assumed SFE. By appropriate adjustment of the SFE, it is possible to reproduce similar metallicities across different SFHs.  

As shown by the solid black line in Fig. \ref{figure_ML}, the stellar mass-to-light ratio ($M_{\rm pop}/L_{[3.6]}$) increases with galaxy mass due to the greater fraction of stellar remnants and the bottom-heavy gwIMF. For comparison, we have also plotted the case of the IGIMF with constant metallicity, based on the MZR (Eq. \ref{eq:MZrel}, labeled as "IGIMF, MZR" in Fig. \ref{figure_ML}). This model yields lower $M_{\rm pop}/L_{[3.6]}$ ratios compared to the model with evolving metallicity. This is because the metallicity of galaxies modeled with a constant metallicity based on the MZR is higher than the average metallicity in models with evolving metallicity. As a result, these galaxies appear redder (i.e., have higher $L_[3.6]$) and produce less massive stellar remnants, both factors contributing to lower $M_{\rm pop}/L_{[3.6]}$ ratios in the constant metallicity models.

Additionally, we performed a set of calculations incorporating an IGIMF model with an evolving metallicity  where star formation declines gradually with a characteristic timescale of $\tau = 10$ Gyr. This new model suggests that a decreasing SFR leads to a higher $M_{\rm pop}/L_{[3.6]}$ ratio which is shown by the solid green line in Fig. \ref{figure_ML}. The impact of a declining SFR is particularly pronounced in high-mass galaxies, as their gwIMF is more top-heavy in the early stages of galaxy formation, resulting in a larger fraction of stellar remnants.

\begin{figure}
    \includegraphics[width=\columnwidth]{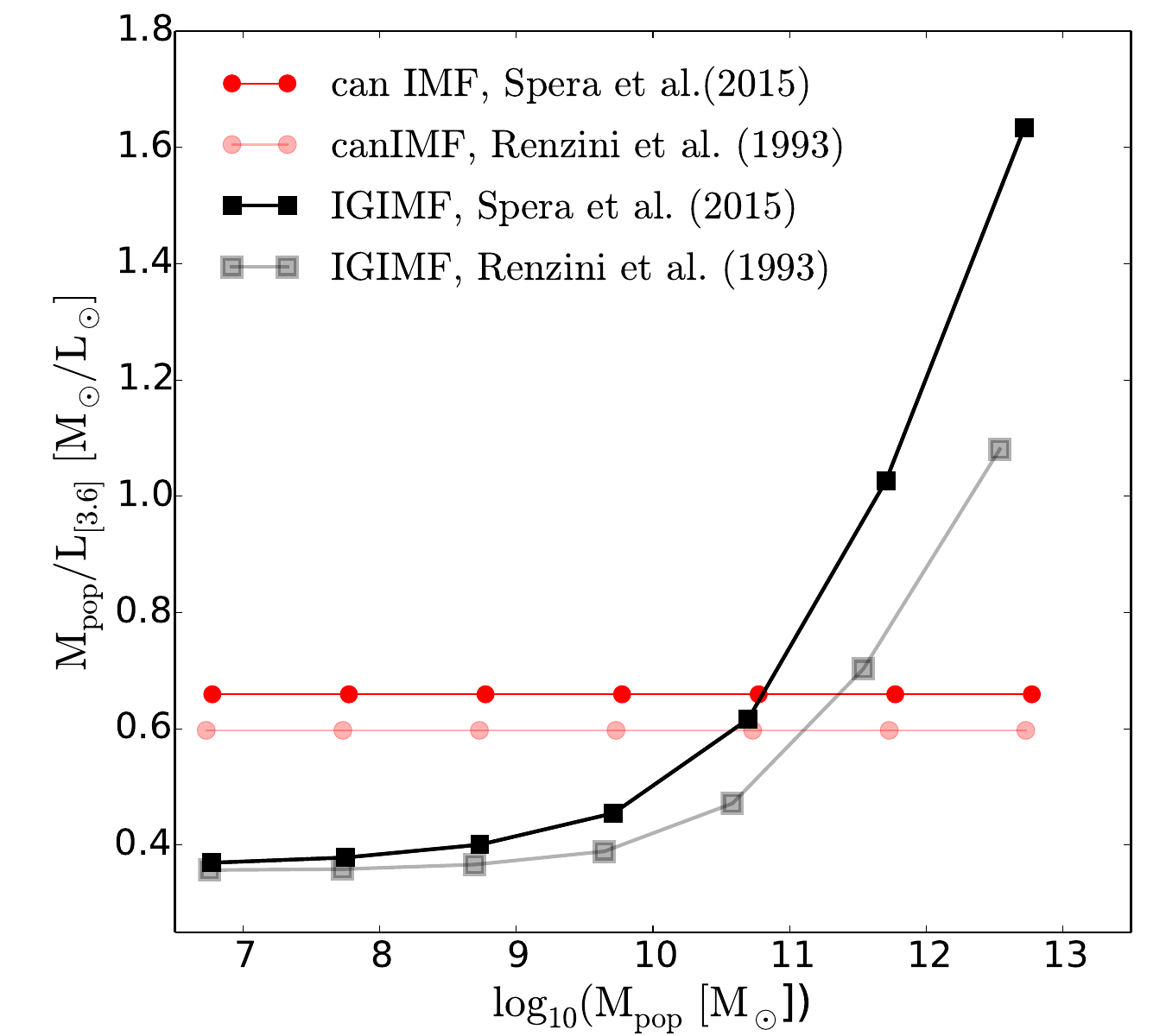}
        \caption{ $M_{\rm pop}/L_{[3.6]}$ in dependence on the present-day stellar mass assuming a constant SFR and time-evolving metallicities, but with different stellar remnant prescriptions. Circle symbols represent models with an invariant canonical IMF, while the square symbols correspond to the IGIMF case. The black squares and filled red circles are identical to Fig. \ref{figure_ML}. The dark lines correspond to models using the metallicity-dependent remnants prescriptions of \citet{Spera2015}, while the light lines illustrate models with the prescriptions of \citet{Renzini93}. }
    \label{figure_ML2}
\end{figure}

Figure \ref{figure_ML2} illustrates the dependence of $M_{\rm pop}/L_{[3.6]}$   on the present-day stellar mass, assuming a constant SFR and time-evolving metallicities, but with different prescriptions for stellar remnants. As shown, adopting the metallicity-dependent prescription for stellar remnants from \citet{Spera2015} leads to higher $M_{\rm pop}/L_{[3.6]}$ ratios, particularly for high-mass galaxies in the IGIMF context. This is because high-mass galaxies, which have lower metallicities in their early stages, yield a greater mass of stellar remnants.

We note that our chemical evolution model adopts the instantaneous recycling approximation within a closed-box framework and therefore does not explicitly include the delayed iron enrichment from SNe Ia. Although SNe Ia contributes significantly to iron enrichment on Gyr timescales, they only contribute approximately 10\% of the total metallic mass of a galaxy \citep{Peeples2014, Yates2024}. For this reason, we consider the omission of SNe Ia under the instantaneous recycling approximation to be a reasonable simplification for the scope and focus of our study.

\begin{figure}
     \includegraphics[width=\columnwidth]{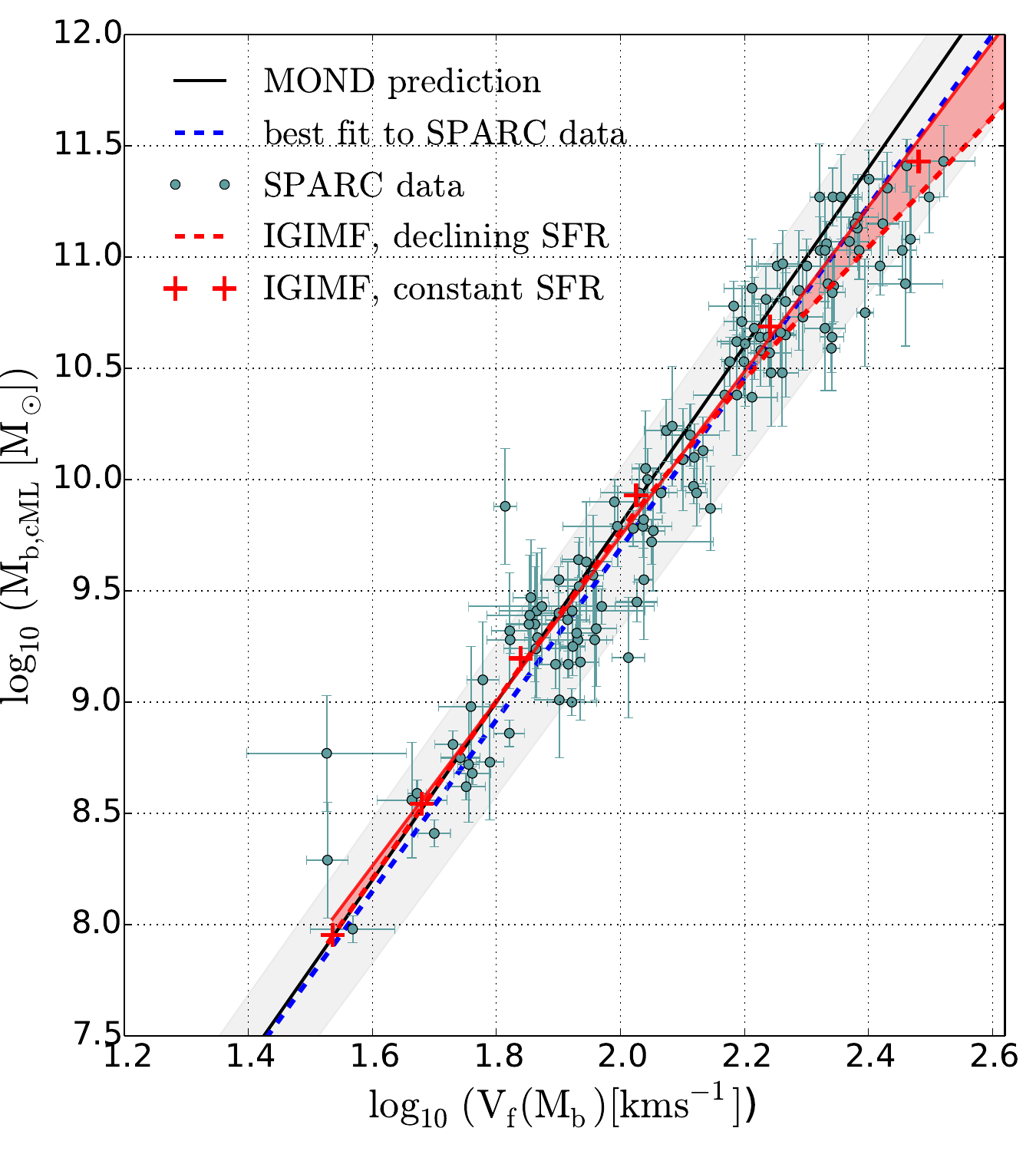}
    \caption{Effect of a varying stellar IMF on the BTFR for different SFHs. The black solid line represents the MOND prediction relating the baryonic mass of galaxies to their flat rotation velocity (Eq. \ref{eq:BTFR0}). The blue dashed line (Eq. \ref{eq:tau_star}) shows the slope of the BTFR derived from the SPARC sample by \citet{Lelli2016}, assuming an invariant gwIMF, which deviates from the MOND prediction. The $M_{b,cML}$ refers to the  derived baryonic mass by assuming a constant \( M_{\rm pop}/L_{\rm [3.6]} = 0.5\). Assuming MOND is correct and that the gwIMF follows the IGIMF theory, an observer would underestimate the true baryonic mass, particularly in high-mass galaxies, if adopting a fixed \( M_{\rm pop}/L_{\rm [3.6]} = 0.5\).  The red crosses with the corresponding best-fitted solid red line illustrate the deviation between the predicted and actual mass when adopting a fixed mass-to-light ratio of $M_{\rm pop}/L_{\rm [3.6]} = 0.5$, instead of using the true $M_{\rm pop}/L_{\rm [3.6]}$ values in the IGIMF theory that are larger for a given $V_{\rm f}\geq 100 ~\rm km/s $. } The impact of a declining SFR is shown by the red dashed line, demonstrating how it leads to increasing discrepancies from the true baryonic mass when the fixed $M_{\rm pop}/L_{\rm [3.6]} = 0.5$ is assumed. This deviation from the MOND prediction is shown by the red shaded area, with the upper boundary corresponding to a constant SFR and the lower boundary corresponding to a slowly declining SFR ($\tau=10 ~\rm Gyr$). 
    \label{figure_TF}
\end{figure}

%Note that, according to MOND predictions, $M_{\rm b,app}$ is equal to the apparent baryonic mass, $M_{\rm b}$

\section{The Tully-Fisher relation in the IGIMF context}
\label{sec:BTFR}

The Tully–Fisher Relation (TFR) is an empirical correlation observed for spiral galaxies, linking their intrinsic baryonic mass to their rotational velocity. This relationship serves as a tool for estimating galactic intrinsic luminosity and distances and offers insights into galaxy formation and dynamics. \citet{Lelli2019} conducted a comprehensive study of the Baryonic Tully–Fisher Relation (BTFR) using 153 galaxies from the SPARC sample. They explored various definitions of characteristic rotational velocities derived from HI and $H_\alpha$ rotation curves, as well as HI line widths from single-dish observations. Their findings indicated that the tightest BTFR is achieved when using the mean velocity along the flat portion of the rotation curve. The baryonic mass of each galaxy is estimated as
\begin{eqnarray}
    \centering
   M_{ \rm b} = M_{\rm g}+\frac{M_{\rm pop}}{L_{[3.6]}}\times L_{[3.6]} , 
    \label{eq:tau_star1}
\end{eqnarray}
where $ L_{[3.6]}$ is the luminosity in the $[3.6]$ micron band. The mass-to-light ratio is assumed to be constant, $ M_{\rm pop}/L_{[3.6]} = 0.5$, for all galaxies from low to very high mass galaxies, as suggested by stellar population synthesis models based on an invariant canonical IMF assumption. By performing a linear fit of the form
\begin{eqnarray}
    \centering
    \log_{10}\left(\frac{M_{\mathrm{b}}}{\rm{M_{\odot}}}\right) =A + B \times \log_{10}\left(\frac{V_{\mathrm{f}}}{\rm{km \, s^{-1}}}\right) \, , 
    \label{eq:tau_star}
\end{eqnarray}
to the SPARC data, $ A=1.99\pm 0.18$ and slope $B=3.85 \pm 0.09$, which is shown by the blue dashed line in Figure \ref{figure_TF}.

In MOND, the Baryonic Tully-Fisher Relation (BTFR) naturally emerges as a fundamental prediction, with a fixed slope of 4 (Eq. \ref{eq:BTFR0}). 

The empirical findings of \citet{Lelli2019} generally align with MOND's predictions, as illustrated by the solid black line in Figure \ref{figure_TF}, indicating that MOND can successfully reproduce the observed BTFR. However, the best-fit linear regression to the data reveals a deviation from MOND’s expected slope of 4, particularly for high-mass, star-forming (i.e., disk) galaxies. As shown in Figure \ref{figure_TF}, these galaxies exhibit systematically higher velocities than those predicted by MOND, suggesting a potential tension between the theoretical framework and observational data. Alternatively, for a given $V_{\rm f}\geq 100 ~\rm km/s $ the galaxies need to have a larger baryonic mass to be consistent with the BTFR.

However, it is important to note that obtaining the baryonic mass of SPARC galaxies from Eq. \ref{eq:tau_star} relies on a specific assumption that the gwIMF is invariant. Based on this assumption, the luminosity of a galaxy is transformed to a stellar population mass using the constant mass-to-light ratio $M_{\rm pop}/L_{[3.6]} = 0.5$. But if the gwIMF follows the IGIMF theory, the mass-to-light ratio ($M_{\rm pop}/L_{[3.6]}$) of a galaxy would depend on its  SFH and metallicity. As shown in Figure \ref{figure_ML}, high-mass galaxies with larger SFRs tend to have larger $M_{\rm pop}/L_{[3.6]}$ ratios compared to low-mass galaxies in the context of the IGIMF theory.

To accurately determine the baryonic mass of a galaxy from Eq. \ref{eq:tau_star1}, one must apply the correct $M_{\rm pop}/L_{\rm [3.6]}$ ratio of the stellar population, including remnants,
\begin{equation}
    \frac{M_{\rm pop}}{L_{\rm [3.6]}} =  \frac{M_{*} + M_{\mathrm{rem}}}{L_{[3.6]}}. \, 
\label{eq:ML}
\end{equation}
Note that the observed quantities are $L_{\rm 3.6}$ and $V_{\rm f}$, and $M_{\rm tot} > M_* + M_{\rm rem}$ always. This model galaxy has a luminosity, $L_{\rm [3.6]}$, which, when transformed into an {\it apparent stellar population} with $M_{\rm pop}/L_{\rm [3.6]} = 0.5$, will correspond to a different apparent total baryonic mass,
\begin{equation}
     M_{\rm b, cML}= 0.5 \, L_{[\rm 3.6]} + M_{\rm g}(L_{\rm [3.6]}).
    \label{eq:Mbapp}
\end{equation}
Assuming MOND is correct, all galaxies should lie on the BTFR predicted by MOND. However, if the analysis of an observed galaxy is performed assuming the gwIMF is invariant such that the galaxy’s luminosity, $L_{[3.6]}$, is transformed to a stellar population mass using the constant mass-to-light ratio, $ M/L_{[3.6]} = 0.5$, the resulting baryonic mass $M_{\rm b, cML}$ will place the galaxy offset from the BTFR.

\subsection{Effect of the SFR on the BTFR}

To investigate the effect of the varying gwIMF on the BTFR, we calculate a grid of galaxies with different total masses within the framework of the IGIMF theory. We calculate the present-day luminosity, the $M_{\rm pop}/L_{[3.6]}$ ratio, and the baryonic mass for each galaxy assuming two different star formation histories: a constant and a declining SFR.

\subsubsection{Constant SFR}

%Assuming a constant SFH, we calculate the present-day luminosity and mass-to-light ratio for each galaxy under the IGIMF theory. 

For each galaxy, the corresponding asymptotic $V_{\rm f}$ is determined from its $M_{\rm b}$, according to the MOND prediction for the BTFR (Eq. \ref{eq:BTFR0}). The apparent baryonic mass is also calculated by multiplying the mass-to-light ratio $M_{\rm pop}/L_{\rm [3.6]} = 0.5$ by the luminosity of each galaxy. This leads to different stellar masses for the galaxies compared to their actual values. 

In this case, high-mass galaxies will have their stellar mass underestimated, while low-mass galaxies will have their stellar mass overestimated by an observer using the canonical gwIMF. However, it is important to note that low-mass galaxies are gas-dominated, so this effect is negligible for their baryonic mass.

The impact of a varying IMF on the functional form of the BTFR is illustrated  in Fig.~\ref{figure_TF}. The resulting offsets for different galaxies are represented by red cross symbols, with the solid red line indicating the best-fit to these points. These cross symbols show the baryonic masses as analyzed by the observers. These deviate from the true BTFR when the gwIMF varies with the SFR and the metallicity according to IGIMF theory, assuming a constant SFR over the galaxy's evolution. As can be seen in Fig. \ref{figure_TF}, the red solid line is nearly exactly in agreement with the dashed blue line given by Eq. \ref{eq:tau_star} as fitted by \citet{Lelli2019}.

%and the shaded red region  The upper boundary of this region 

\subsubsection{Declining SFR}

Next, we explore the effect of a slowly declining SFR with a timescale of $\tau = 10$ Gyr, on the mass-to-light ratio and baryonic mass.   This results in a larger deviation from the MOND-predicted BTFR compared to the constant SFR scenario, particularly for high-mass galaxies, which is shown as the dashed red line in Fig. \ref{figure_TF}. As the SFR declines, the actual mass-to-light ratio of high-mass galaxies increases above the constant value of 0.5 in the 3.6 $\mu$m  band, leading to larger offsets. The shaded red region illustrates the range of baryonic masses that would be inferred erroneously assuming the gwIMF is the invariant canIMF if galaxies had SFHs varying between constant and slowly declining SFRs.

Adopting a faster-declining SFR leads to an even larger deviation, since the actual $M_{\rm pop}/L_{\rm [3.6]}$ ratio of high-mass galaxies would be higher than 0.5. These results suggest that if MOND is correct and galaxies follow the MOND-predicted BTFR, the observed deviation from this relation in high-mass galaxies is a natural consequence of the IGIMF theory.

The IGIMF framework indeed predicts a significant variation in the stellar M/L ratios across galaxies, particularly due to its sensitivity to the SFR and SFH. However, this variation does not necessarily lead to a large deviation from the BTFR, especially in low-mass galaxies, where the gas component dominates the baryonic mass. In such systems, the contribution of stellar mass and hence the M/L ratio have a relatively minor impact on the total baryonic budget. In contrast, for more massive galaxies, where the stellar mass comprises a larger fraction of the total baryonic mass, variations in the M/L ratio play a more critical role. This is particularly relevant when different SFHs are considered, as they affect M/L even under an invariant gwIMF. Under the IGIMF framework, which inherently links the IMF shape to the SFR, these effects are amplified, leading to a potentially larger scatter in M/L ratios.  This tension can be eased by viewing the IGIMF framework as a way to better understand and constrain galaxy evolution. The scatter introduced by IGIMF-driven M/L variations could help constrain the SFH across different galaxies. In this context, the small observed scatter in the BTFR would imply that galaxies follow relatively constrained SFHs such as constant SFRs over their lifetimes, as is independently indeed suggested by the $\approx1000$ galaxies in the local cosmological volume \citep{Kroupa+20b}.

\section{Conclusions}
\label{sec:Conclusion}

In this study, we explore the impact of an environment-dependent gwIMF on the BTFR, focusing on how systematic differences in stellar populations lead to apparent deviations from the expected relationship. By considering the IGIMF theory to quantify the gwIMF in different galaxies, we demonstrate that variations in the mass-to-light ratio due to metallicity and SFH can lead to discrepancies between the true and observed baryonic masses of galaxies.
Specifically, the luminosity and stellar remnant masses of galaxies depend significantly on the SFH and metallicity, which in turn influence the galaxy's mass-to-light ratio and, consequently, the inferred baryonic mass.

Our results indicate that for high-mass galaxies, assuming a constant mass-to-light ratio leads to an underestimation of their stellar population masses, whereas for low-mass galaxies, the stellar population masses are overestimated, assuming the IGIMF theory is correct. 
However, it should be noted that the baryonic mass of low-mass galaxies is less affected by changes in the gwIMF due to their gas-dominated nature. The systematic deviations from the true BTFR are most pronounced in massive disk galaxies where the SFH has declined slowly over time.

Furthermore, we find that the IGIMF theory can explain the observed offsets in the BTFR, particularly in high-mass galaxies, which exhibit higher rotational velocities than predicted by MOND. This suggests that the deviations observed in the BTFR for these galaxies may be a direct consequence of the varying mass-to-light ratios predicted by the IGIMF theory. These galaxies have a larger relative contribution by stellar remnants and faint low-mass stars to their baryonic masses because of the top-heavy and bottom-heavy gwIMF, which naturally results in the IGIMF theory. Consequently, when observed under the assumption of an invariant gwIMF, the apparent baryonic masses of galaxies deviate from the true MOND-predicted relation.  This highlights the importance of incorporating variations in stellar population models driven by a metallicity and SFH-dependent gwIMF. However, we emphasize that any assumed variation of the gwIMF must conform with the large body of empirical constraints available from different stellar populations \citep{Kroupa2024b}, as is indeed fulfilled by IGIMF theory.  We also emphasize that the IGIMF theory and its quantification have been developed over the past dozen years independently of any MOND application.

\section*{Data availability} 

The data underlying this article are available in the article.

\section*{Acknowledgements}
AHZ acknowledges support from the Alexander von Humboldt Foundation.
HH is grateful to the staff at the Helmholtz-Institut für Strahlen- und Kernphysik (HISKP) and Argelande Institute for Astronomy (AIfA) for their hospitality and acknowledges financial support from the SPYDOR group at the University of Bonn. PK thanks the DAAD Bonn-Prague Eastern European Exchange Program. TJ acknowledges the MUNIAward in Science and Humanities MUNI/I/1762/2023.

\bibliographystyle{mnras}
\bibliography{reference}

\bsp
\label{lastpage}
\end{document}